\documentclass[manuscript]{aastex61}
\makeatletter
\let\@dates\relax
\makeatother

\usepackage{subfigure}
\usepackage{natbib}
\usepackage{graphicx}
\usepackage{multirow}
\usepackage{epstopdf}
\usepackage{amsmath}
\usepackage{url}
\usepackage{hyperref}
\usepackage{amssymb}
\usepackage{booktabs}
\usepackage{lineno}
\usepackage{color}

\shorttitle{GBM Updates for LIGO/Virgo O3}
\shortauthors{Goldstein, et al.}

\begin{document}

\title{Updates to the {\it Fermi}-GBM Targeted Sub-threshold Search in Preparation for the Third Observing Run of LIGO/Virgo}

\author[0000-0002-0587-7042]{Adam~Goldstein}
\affiliation{Science and Technology Institute, Universities Space Research Association, Huntsville, AL 35805, USA}

\author{Rachel~Hamburg}
\affiliation{Space Science Department, University of Alabama in Huntsville, 320 Sparkman Drive, Huntsville, AL 35899, USA}

\author{Joshua~Wood}\thanks{NASA Postdoctoral Fellow}
\affiliation{NASA Marshall Space Flight Center, NSSTC, 320 Sparkman Drive, Huntsville, AL 35805, USA}

\author{C.~Michelle~Hui}
\affiliation{Astrophysics Office, ST12, NASA/Marshall Space Flight Center, Huntsville, AL 35812, USA}

\author{William H. Cleveland}
\affiliation{Science and Technology Institute, Universities Space Research Association, Huntsville, AL 35805, USA}

\author{Daniel~Kocevski}
\affiliation{Astrophysics Office, ST12, NASA/Marshall Space Flight Center, Huntsville, AL 35812, USA}

\author{Tyson~Littenberg}
\affiliation{Astrophysics Office, ST12, NASA/Marshall Space Flight Center, Huntsville, AL 35812, USA}

\author{Eric~Burns}
\affiliation{NASA/Goddard Space Flight Center, Greenbelt, MD 20771, USA}\thanks{NASA Postdoctoral Fellow}

\author[0000-0001-5078-9044]{Tito~Dal~Canton}
\affiliation{Max-Planck-Institut f\"ur Gravitationsphysik (Albert Einstein Institut), Am M\"uhlenberg 1, D-14476 Potsdam-Golm, Germany}
\affiliation{LAL, Univ. Paris-Sud, CNRS/IN2P3, Universit Paris-Saclay, F-91898 Orsay, France}

\author[0000-0002-2149-9846]{P{\'e}ter~Veres}
\affiliation{Center for Space Plasma and Aeronomic Research, University of Alabama in Huntsville, 320 Sparkman Drive, Huntsville, AL 35899, USA}

\author{Bagrat~Mailyan}
\affiliation{Center for Space Plasma and Aeronomic Research, University of Alabama in Huntsville, 320 Sparkman Drive, Huntsville, AL 35899, USA}

\author[0000-0002-0380-0041]{Christian Malacaria}\thanks{NASA Postdoctoral Fellow}
\affiliation{NASA Marshall Space Flight Center, NSSTC, 320 Sparkman Drive, Huntsville, AL 35805, USA}
\affiliation{Universities Space Research Association, NSSTC, 320 Sparkman Drive, Huntsville, AL 35805, USA}

\author{Michael~S.~Briggs}
\affiliation{Center for Space Plasma and Aeronomic Research, University of Alabama in Huntsville, 320 Sparkman Drive, Huntsville, AL 35899, USA}
\affiliation{Space Science Department, University of Alabama in Huntsville, 320 Sparkman Drive, Huntsville, AL 35899, USA}

\author[0000-0002-8585-0084]{Colleen~A.~Wilson-Hodge}
\affiliation{Astrophysics Office, ST12, NASA/Marshall Space Flight Center, Huntsville, AL 35812, USA}

\begin{abstract}
In this document, we detail the improvements made to the Fermi GBM targeted sub-threshold search for counterparts to LIGO/Virgo gravitational-wave triggers. We describe the implemented changes and compare the sensitivity of the O3 search to that of the version of the search that operated during O2.  Overall, we have improved both the sensitivity and speed of the targeted search.  Further improvements to the search have been made for the O3b observing run, including automated upper limits estimation and incorporating the updated localization systematic with the new version of the search.
\end{abstract}

\section{Introduction}
The {\it Fermi} Gamma-ray Space Telescope's Gamma-ray Burst Monitor (GBM) is currently the most prolific detector of 
Gamma-ray Bursts (GRBs), including short-duration GRBs.  The GBM triggers on $\sim$ 240 GRBs per year, $\sim$40 of which 
are short GRBs; localizes GRBs to an accuracy of a few degrees; has a broad energy band (8 keV--40 MeV) at high spectral 
resolution for spectroscopy; and records data at high temporal resolution (down to $\sim$2 $\mu$s)~\citep{Meegan09}.  Recently 
the detection rate of short GRBs has been increased with a ground-based `untargeted' search to detect fainter events which did not trigger 
GBM\footnote{\url{https://gcn.gsfc.nasa.gov/admin/fermi_gbm_subthreshold_announce.txt}}.  The detection of GRBs by GBM have led to a 
plethora of analyses, including joint analyses with the {\it Fermi} Large Area Telescope, {\it Swift}, and ground-based optical and 
radio telescopes.  Although the localization of GRBs by GBM is rough in comparison to the capabilities of {\it Swift}, collaborative efforts have enabled wide-field optical telescopes to tile the large GBM localization regions and discover the GRB optical afterglow independent of any other gamma-ray instrument~\citep{Singer15, Lipunov15}.  The next generation of wide-field telescopes now coming online are even more capable~\citep[e.g.][]{Coughlin19}.

Motivated by the possibility that short GRBs are caused by compact binary mergers that produce gravitational 
waves and may be observable by LIGO, \citet[][hereafter LB15]{Blackburn15} developed a method to search the GBM continuous 
data for transient events in temporal coincidence with a LIGO compact binary coalescence trigger.  The LB15 search operated by 
ingesting a LIGO trigger time and optionally a LIGO localization probability map and searched for a signal over different 
timescales around the LIGO trigger time. The search looked for a coherent signal in all 14 GBM detectors by using spectral 
templates that are convolved with the GBM detector responses calculated over the entire un-occulted sky to produce an 
expected count rate signal in each detector.  The expected counts in each detector were compared to the observed counts, taking 
into account a modeled background component.  A likelihood ratio (LogLR) was then calculated comparing the presence of a signal to the null hypothesis of pure background. LB15 discusses the formalism and implementation of this approach along with an estimation of the detection significance distribution during 2 months of the LIGO S6 run.  This version of the search was also operated by the GBM Team during LIGO O1, resulting in the detection of a weak transient~\citep{Connaughton16,Connaughton18} $\sim$0.4~s after the first gravitational-wave detection, GW150914~\citep{Abbott16}. Additionally, the LB15 search was used to follow up all three LIGO triggers during O1 as well as 56 other sub-threshold GW triggers during O1, presenting the first joint sub-threshold search for coincident GW and EM signals in GBM~\citep{O1Paper}. Because this search is not blind and probes GBM data around a time of interest, we call this the `targeted' search throughout.

During the break between O1 and O2, improvements were made to the targeted search to enhance its sensitivity, including improved background estimation, a replacement of the `hard' spectral template, and utilization of GBM's continuous event data.  These improvements, and the corresponding validation with both background and known real signals, are detailed in~\citet{O2Updates}.  Additionally,~\citet{Kocevski18} utilized the O2 version of the targeted search to show that it can be used to uncover {\it Swift}-detected GRBs that were observed by GBM but too weak to trigger the onboard algorithms.  Although the targeted search was not required for the detection of the groundbreaking discovery of GW170817/GRB 170817A~\citep{MMA}, it was used to perform a more accurate GBM localization~\citep{Goldstein17} and to determine the maximal distance at which GBM could have detected the GRB~\citep{GW-GRB}. The targeted search was also used to study the different spectral components present in GRB~170817A, leading to the search for other GRBs with similar signatures~\citep{150101B, vonKienlin19}.

Now, during the third observing run of LIGO/Virgo, we present further updates to the targeted search, improving both its speed and sensitivity.  This version of the search will be operated during O3b, and we detail the improvements and validation of those improvements in the following sections.

\clearpage

\section{Changes to the Targeted Search\label{sec:Changes}}

\subsection{Background Estimation\label{sec:BackEst}}
As discussed in~\citet{O2Updates}, the background is estimated using an un-binned Poisson maximum likelihood technique with a fixed duration sliding window of length $T$. The length of this window was chosen to be 125 s based on the fact that this timescale accurately modeled backgrounds during testing with several days of GBM data. We noticed, however, that this method can produce a poor background model in cases where the background rate increases or decreases more rapidly than normal, particularly during approach and exit from the SAA. Most of these cases are handled by restricting the search region to avoid time periods close to SAA but during O3a we observed $\sim$1\% of searches to be affected by poor background fits due to rate changes on timescales shorter than 125 s. To handle these edge cases, we implemented a check for goodness of fit within two 30 s duration windows extending directly before and after the search range using the existing $\chi^2$ criterion described in~\citet{O2Updates}. The check examines all background points within the two windows and determines whether the fraction of bad background fit points exceeds 30\% for any single NaI detector over the energy range 11 keV to 975 keV. If this is true, the entire background fit is redone with a 50\% smaller sliding time window and again checked for goodness of fit. This procedure continues until the fraction of bad background fit points is less than 30\% for all NaI detectors or until it has gone through two iterations. This was shown to improve background fits in all problem cases during testing on all O3a events.

\subsection{Atmospheric Scattering\label{sec:AtmoScat}}
The energy responses for the GBM detectors are angular-dependent and have contributions from two primary components: 1) direct incidence and scattering of gamma-ray photons on {\it Fermi} and 2) back-scattering of high-energy photons incident on the Earth's atmosphere into the GBM detectors.  The atmospheric scattering component is more complicated than the direct component because it is dependent on the Earth-detector-source geometry as well as the inherent spectrum of the emitting source.  Generally, the flux from atmospheric scattering is stronger with a harder spectrum and can be the dominant source of flux for detectors depending on the observing geometry.  The atmospheric scattering component is important to consider when localizing sources because an improper consideration of atmospheric scattering flux can significantly bias a source localization away from the true position by attributing too much or too little of the observed flux to atmospheric scattering.  Historically, the GBM localization algorithm considers only flux in the 50-300 keV range for each detector, the energy range over which most of the flux is emitted for most GRBs.  Since the generation of the atmospheric scattering component of the spectral templates is non-trivial and extremely time consuming, the targeted search originally only considered the atmospheric scattering in this energy range, while using the direct flux component for all energy channels.

\begin{figure}
	\begin{center}
		\includegraphics[scale=0.8]{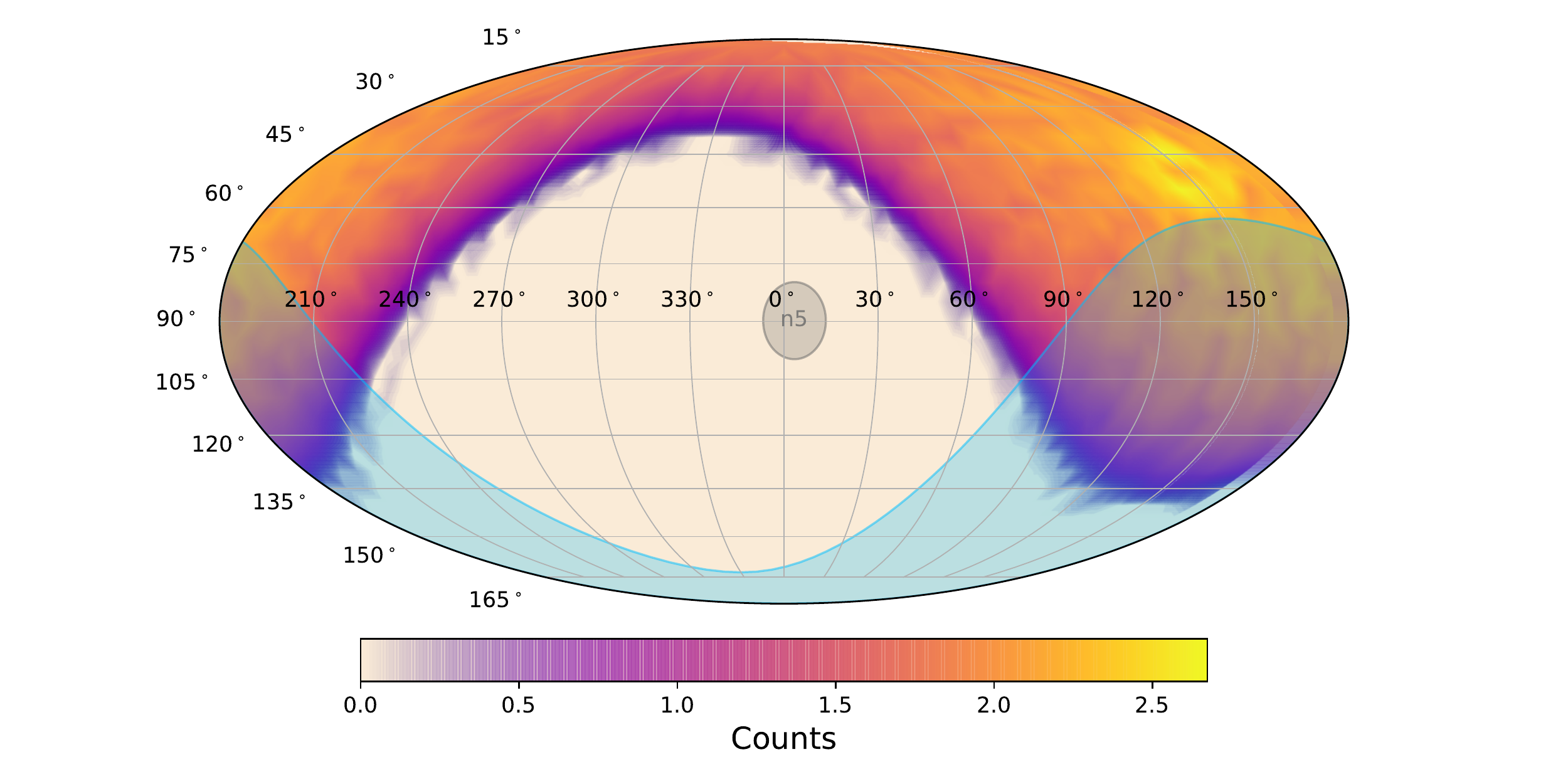}\\
		\includegraphics[scale=0.8]{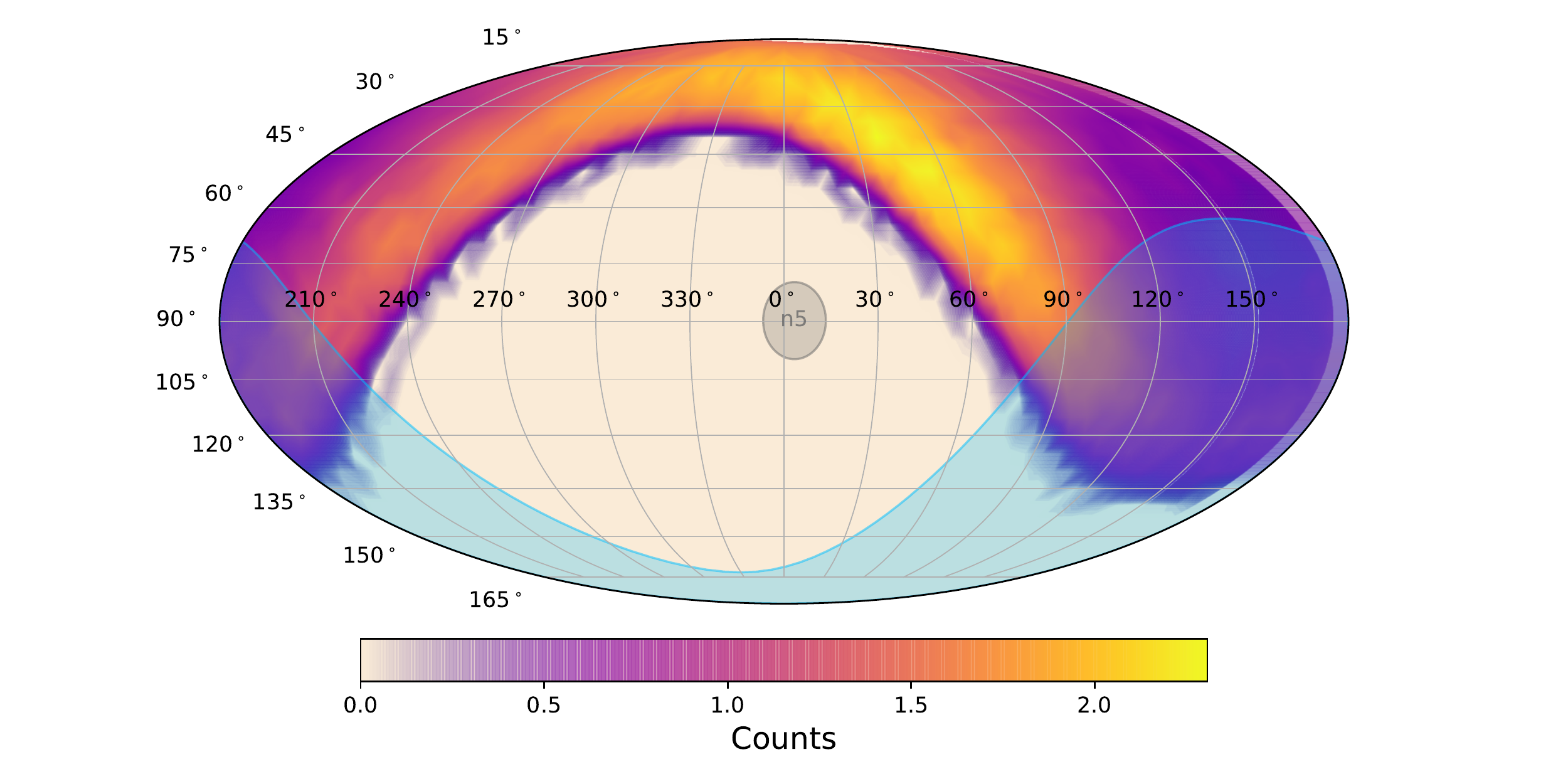}
	\end{center}
\caption{An example of the atmospheric scattering flux from different GRB source positions in the 12--27 keV range (top) and the 290--540 keV range (bottom) for the hard template in detector n5.  Coordinates are in the spacecraft inertial frame where zenith=0 is the pointing of the Fermi LAT. The pointing of detector n5 is shown for context.  The atmospheric scattering component in this example uses a geocenter location of azimuth=150, zenith=130 (Earth shown as the blue shaded area). The flux overlapping the Earth is masked out during the search.
\label{AtmoScat}}
\end{figure}

For O3, we have generated the atmospheric scattering component of the templates for all energy channels in all detectors over the whole sky.  This has also lead to a simplification of the targeted search code when combining the two response components. Shown in Figure~\ref{AtmoScat} are examples of the atmospheric scattering in two different energy channels for the GBM detector n5 for a specific location of the geocenter in spacecraft coordinates. Of note is the fact that the atmospheric scattering contribution can change significantly with energy. Correctly accounting for the atmospheric scattering component in all energy channels is expected to boost the detection statistic for real, coherent signals in the data.\\

\subsection{Temporal Resolution\label{sec:Temporal}}
As discussed in~\citet{O2Updates}, the minimum temporal resolution of the data searched was reduced to 64 ms from 256 ms, but the minimum timescale searched was still 256 ms.  At the time the O2 search was being tested, it was unclear if the 64 ms timescale would lead to a benefit considering the cost of the extra search trials.  Additionally, the computational cost was significant to search down to this timescale, however, the O3 search has considerably improved this situation (see Section~\ref{sec:SkyRes}).
We have further investigated operating the search down to 64 ms to recover shorter signals and have verified that the detection significance of such signals exceeds the cost of the additional trials imposed by this timescale.  Furthermore, to address the issue raised in Section 9 of~\citet{O2Updates} where it was discovered the significance of a signal could change considerably based on the phase of the bins relative to the signal, we have increased the maximum phase shifts of the search from four to eight.  We verified that this does not negatively impact the general post-trials significance of the signals due to the additional trials. 

\subsection{Energy Range\label{sec:EnergyRange}}
\begin{figure}
	\begin{center}
		\includegraphics[scale=0.6]{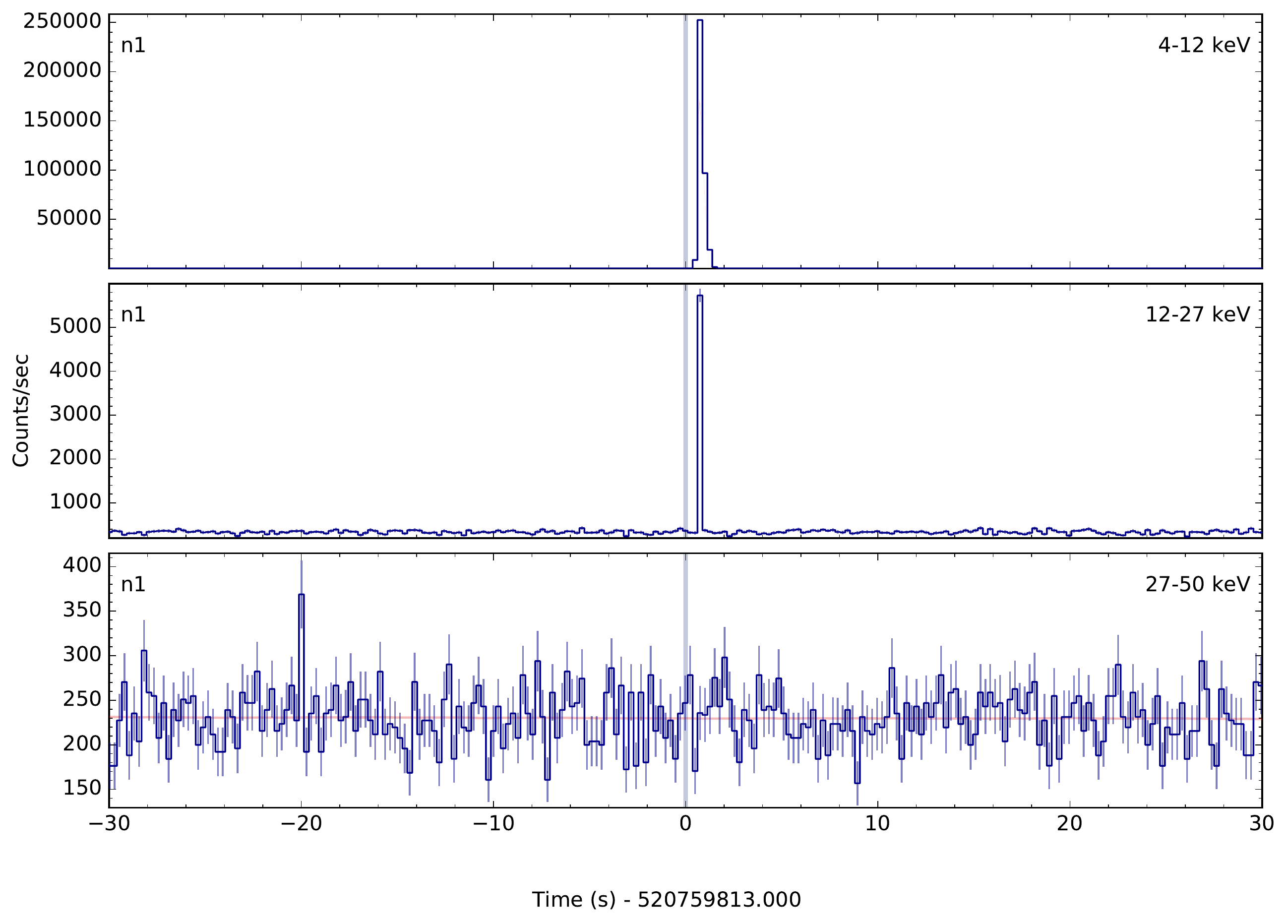}
	\end{center}
\caption{An example of a phosphorescent spike in the lowest energy channels of detector n1. During O2, the lowest energy channel of the NaI detectors was included in the search, resulting in a LogLR $\sim4000$ for this event. For the O3 search, we have removed the lowest channel from consideration, and in this case results in a LogLR of  $\sim14$.
\label{PhosSpike}}
\end{figure}

In previous versions of the targeted search, we utilized all 8 energy channels as defined by the CTIME data (covering $\sim4-1000$ keV).  In many standard GBM analyses, energy channels below 8 keV (the first channel of the NaI CTIME data) are not used.   The detector responses below that energy are poorly modeled, contributing to an increased source of systematic.  Additionally, soft flares from Galactic transients, not thought to be a promising source of detectable transient gravitational-wave signals, tend to populate the lower region of the GBM observing range, as do short, bright phosphorescence spikes from the interactions of energetic cosmic rays in the NaI crystal (an example is shown in Figure~\ref{PhosSpike}).  For these reasons, we have removed the lowest energy channel (4--12 keV) of the NaI detectors from consideration in the targeted search.

Occasionally the search will detect either long flaring emission in the second channel or the background model will be inaccurate due to rapid changes (such as occultation steps, exits from SAA, etc.).  These detections are typically found on the longest timescale of the search, 8.192 s.  This long duration and soft spectrum form an unlikely combination for the detection of a short GRB-like transient, therefore, we are using this combination as a veto during signal searches to automatically filter out likely unrelated signals.

\subsection{Sky Resolution\label{sec:SkyRes}}
Historically, the GBM localization of sources on the sky has been performed on a 1-degree resolution grid on the sky.  This was done because the (expected) localization systematic is on the order of a few degrees. Performing a search for spectrally- and spatially-coherent signals over the entire unocculted sky on such a fine grid is time consuming and is the primary driver for the computation time of the targeted search. Also, since we are searching for weak signals below the GBM triggering threshold, we expect the localization uncertainty to be dominated by statistical uncertainty and not systematic, therefore a 1-degree resolution grid is not necessary.  We have decreased the sky grid resolution to 5 degrees, providing an order of magnitude decrease in run time and have confirmed that this does not significantly impact the sensitivity ($< 0.1$\% change) of the targeted search to detect or localize GRB-like signals.

\subsection{Blackbody Template\label{sec:BlackBody}}
The targeted search is operating with the same three spectral templates that were used in O2: a `soft' Band function, a `normal' Band function, and a `hard' exponentially-cutoff power law. We have added a new blackbody template, with a $kT = 10$ keV, which is motivated by the discovery of the soft thermal-like tail observed in GRB 170817A.  We will not run the targeted search using this fourth template to search for a signal but rather as part of an effort to perform follow-up characterization of a candidate signal.  Another low-redshift GRB, 150101B, was found to have a similar thermal-like tail using this template, thereby lending some evidence that this may be a feature observable in low-redshift short GRBs~\citep{150101B, vonKienlin19}.

\section{Validation}
We have validated the implemented changes by testing the search on above-threshold and below-threshold short GRBs that were also detected by Swift or other instruments.  This sample includes 26 GRBs from~\citet{Kocevski18} as well as 25 sub-threshold GRBs that were found by the blind `untargeted' search and confirmed by other instruments.  The updates to the search have also been tested on random background segments to assess how sensitive the new search is to GRB signals relative to the background.

\subsection{Validation with Signals\label{sec:SignalValidation}}
\begin{figure}
	\begin{center}
		\includegraphics[scale=0.5]{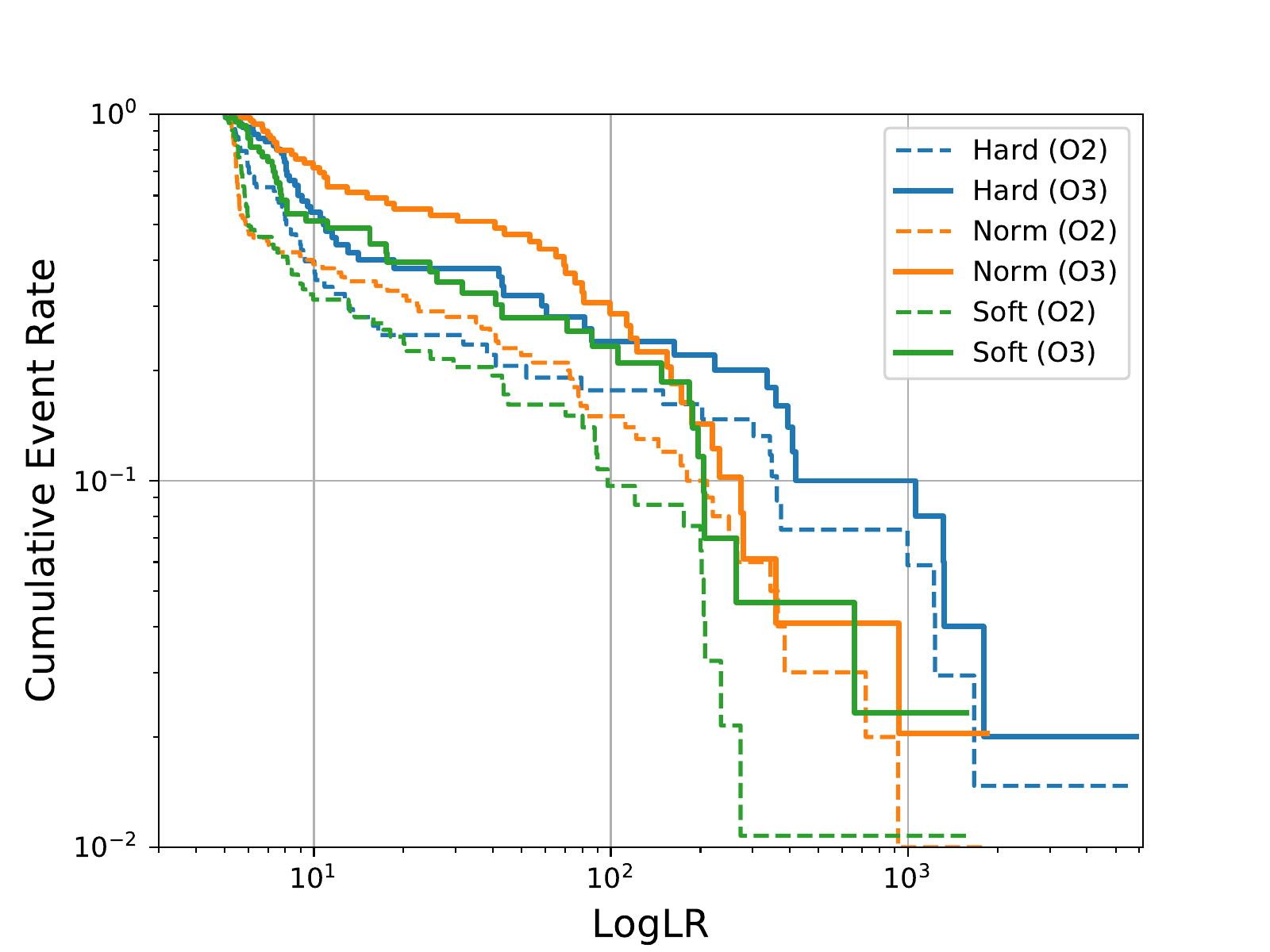}\\
		\includegraphics[scale=0.5]{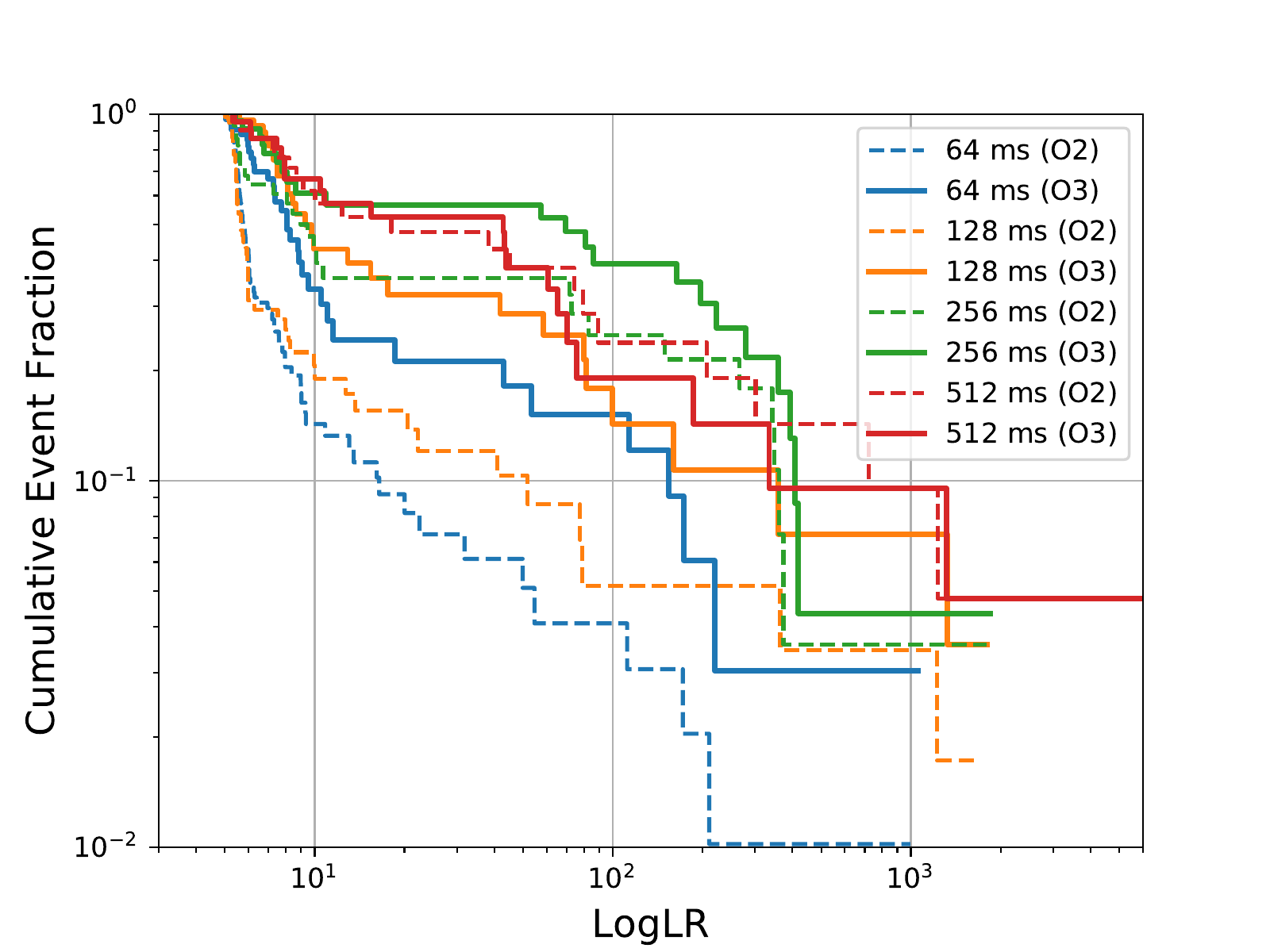}
		\includegraphics[scale=0.5]{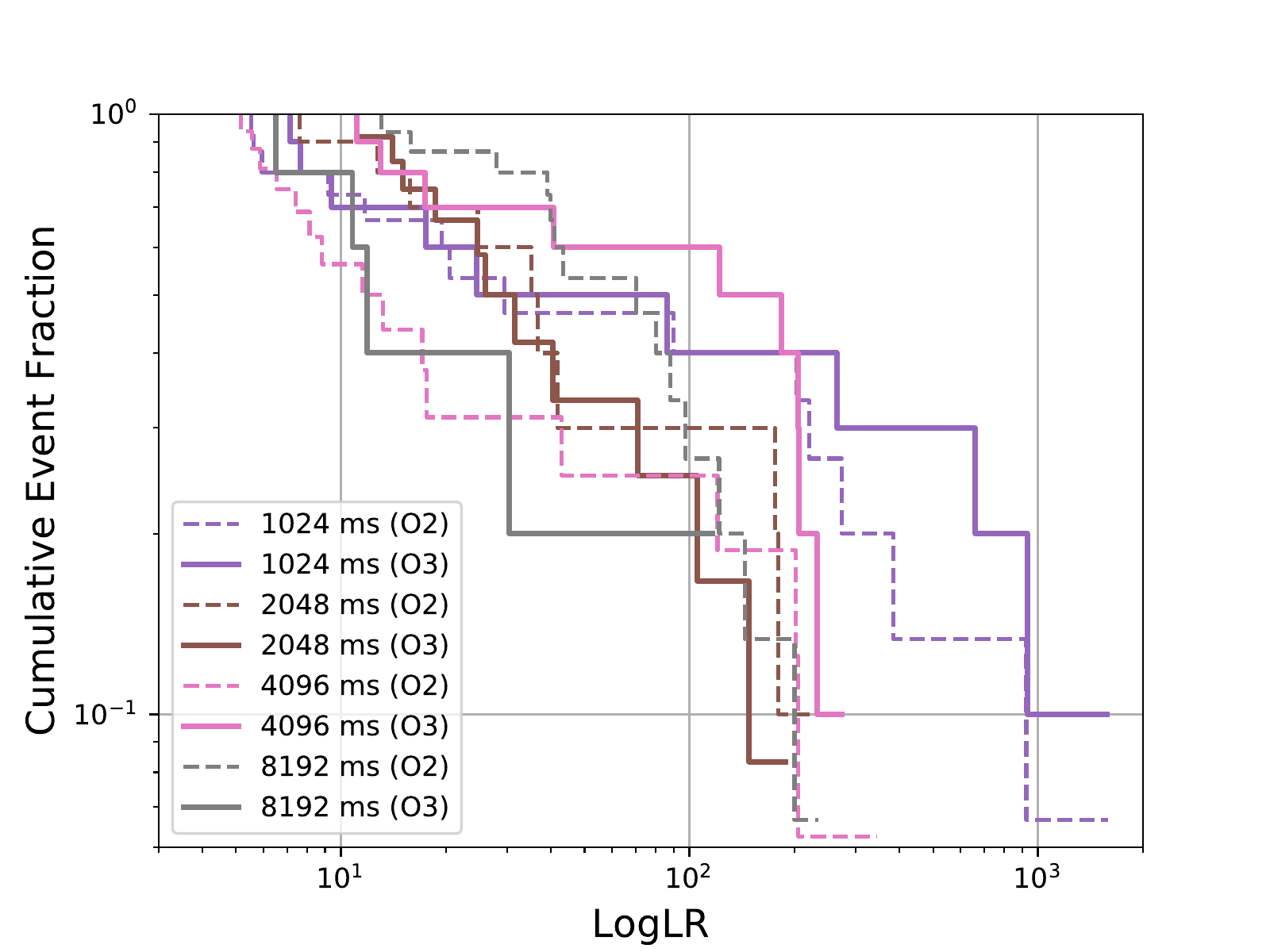}
	\end{center}
\caption{Cumulative distributions of the event rate for true signals compared between the O2 and O3 searches.  The top figure shows distributions of the different spectral templates and the two bottom figures show the distributions of the different timescales searched.
\label{FinalSignal}}
\end{figure}

The sample of 51 confirmed short GRBs were utilized to test each of the changes in Section~\ref{sec:Changes}, however, here we only show the comparison between the final implementations of O2 and O3 searches. Figure~\ref{FinalSignal} shows the comparison of the cumulative event rate between the O2 and O3 searches broken down by the three spectral templates and by the eight search timescales.  Immediately obvious is the improvement (increased LogLR) in the normal and hard spectral templates for the O3 search, which is related to the changes in the atmospheric scattering response described in Section~\ref{sec:AtmoScat}.  Even with the removal of the first energy channel of NaI data, the soft template in the O3 version provides an improvement in the LogLR for real signals. Similarly, the O3 version offers an improvement in the majority of the timescales probed by the search, with five of the eight timescales displaying an improvement spanning nearly the full range of the detection statistic.  Out of the shorter search timescales, only the 512 ms timescale does not show an improvement and is approximately unchanged compared to the O2 search.  The detection statistic on the 2 s timescale is not obviously improved over the full range of LogLR, and tend toward less significant LogLR values for the more significant signals.  Similarly, the 8 s timescale is shifted to lower LogLR values, and can also be mostly explained by the removal of the first energy channel, since softer signals detected by GBM are usually longer in duration.

\subsection{Validation with Background \label{sec:BackgroundValidation}}
\begin{figure}
	\begin{center}
		\includegraphics[scale=0.5]{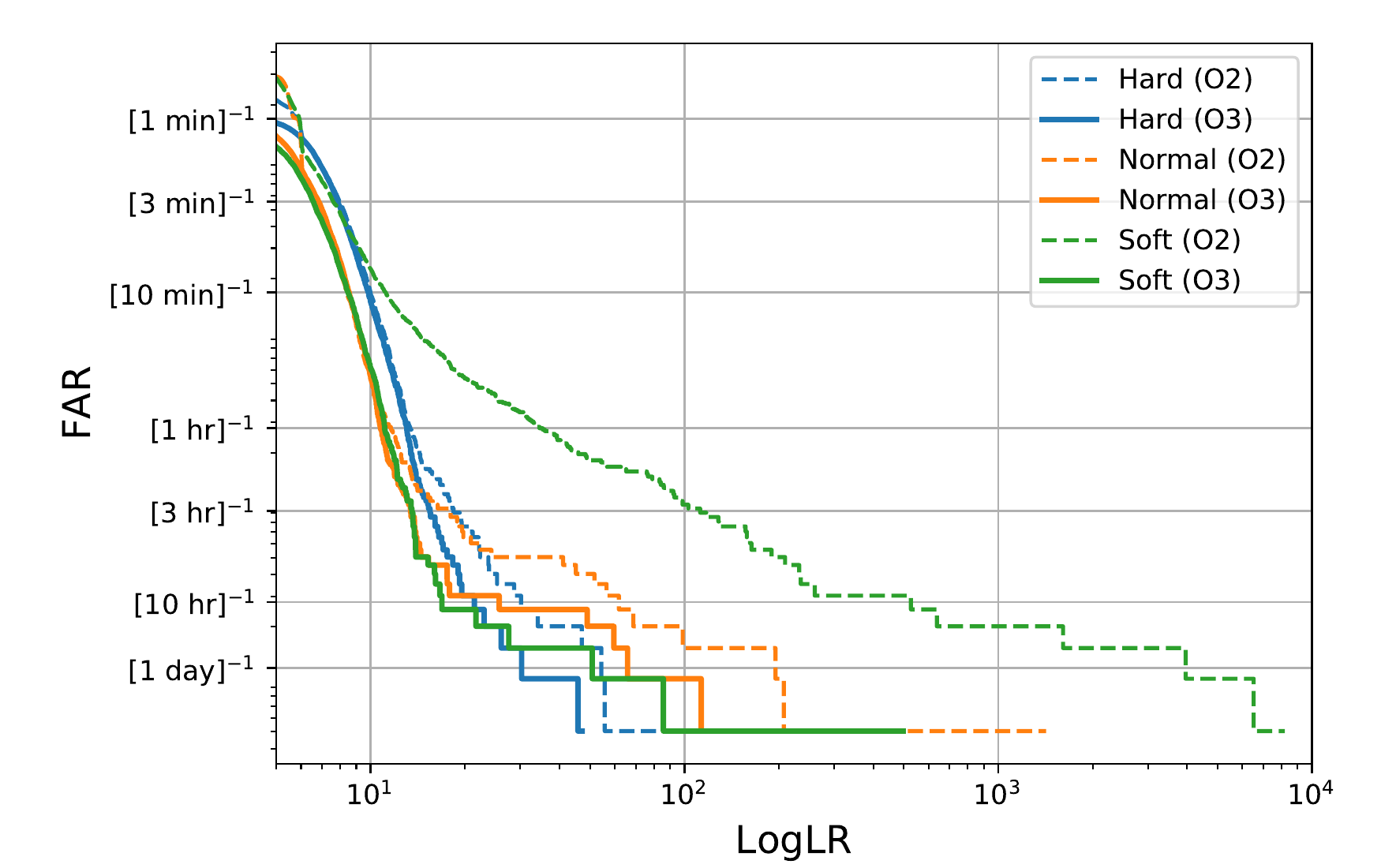}\\
		\includegraphics[scale=0.5]{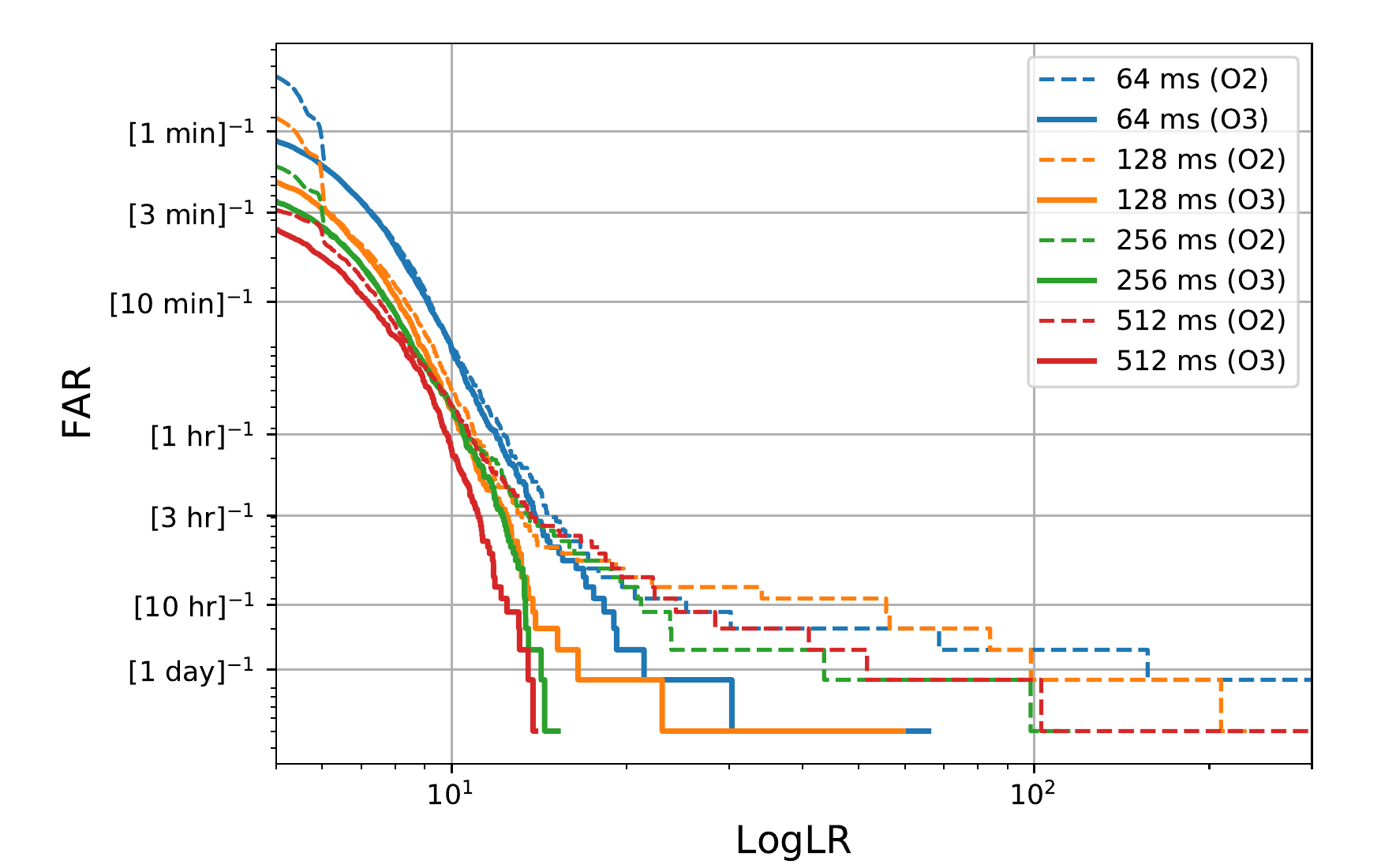}
		\includegraphics[scale=0.5]{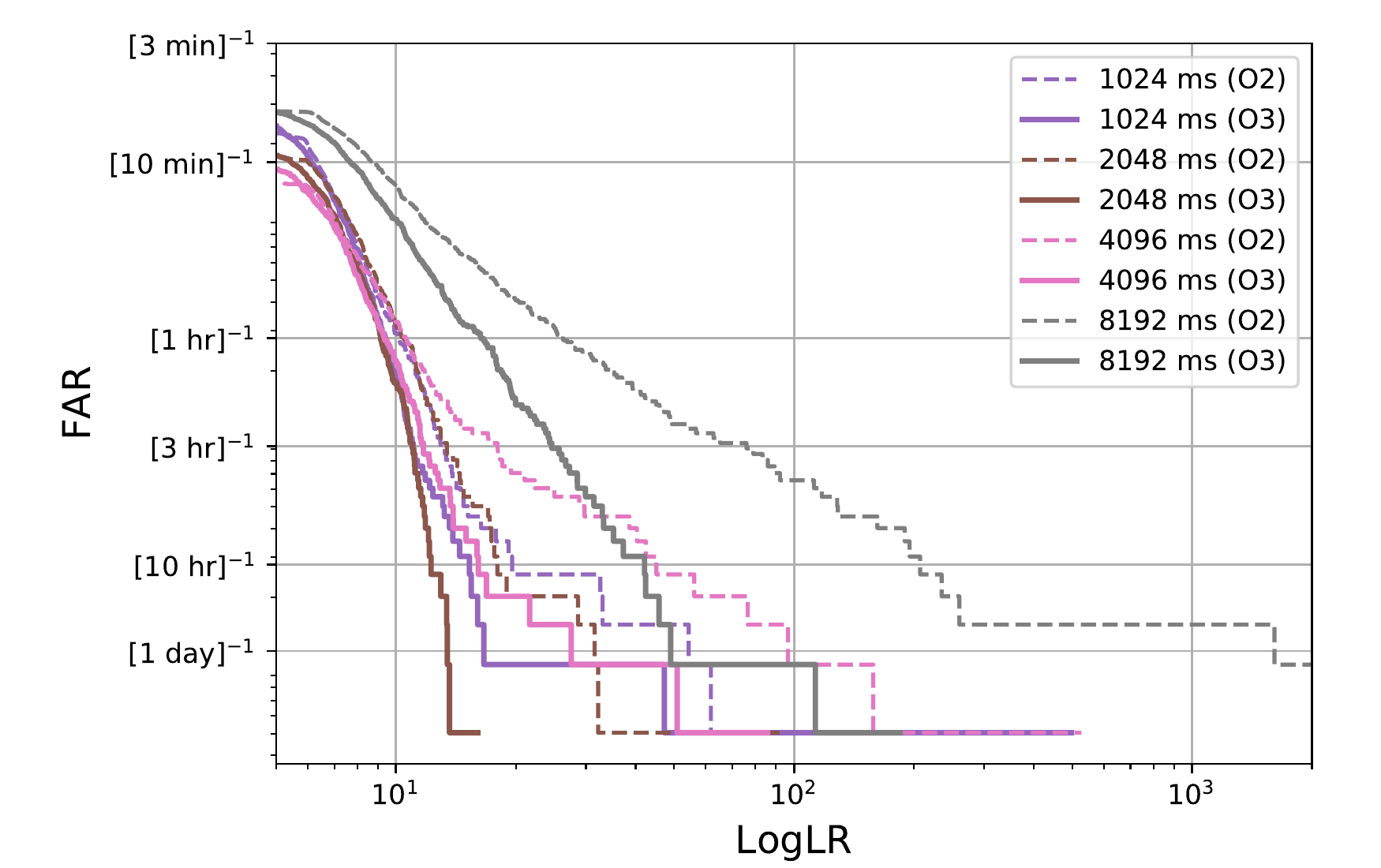}
	\end{center}
\caption{Cumulative distributions of background FAR compared between the
O2 and O3 searches.  The top figure shows the FAR distributions of the different spectral templates and the two bottom figures show the FAR distributions of the different timescales searched.
\label{FinalBack}}
\end{figure}

To evaluate the implemented changes on the background, we ran both the O2 and O3 searches over $\sim$200 ks of GBM data from December~1,~2016 through August~25,~2017, corresponding to random times during O2 where at least two gravitational-wave interferometers were operational. No effort was made to avoid periods where GBM had triggered on a GRB or any other significant signal. Figure~\ref{FinalBack} shows the comparison of the cumulative False Alarm Rate (FAR) between the O2 and O3 searches divided into the spectral templates and timescales.  The most obvious difference between the backgrounds is the significant improvement (decrease) of the FAR for the soft template.  This improvement is mostly due to the removal of the lowest energy channel of the NaI detectors.  Soft flaring sources, local particle activity, and phosphorescent spikes in the lowest energy channels are less significant in the O3 search.  The FAR for the normal and hard templates are improved as well.  Every timescale has an improved FAR, most notable at higher LogLR.  Most of the improvement at the high-LogLR end of the distribution is a result of better treatment of the background around Fermi entrance and exit of the SAA.

\subsection{Overall Comparison\label{sec:Overall}}
\begin{figure}
	\begin{center}
		\includegraphics[scale=0.5]{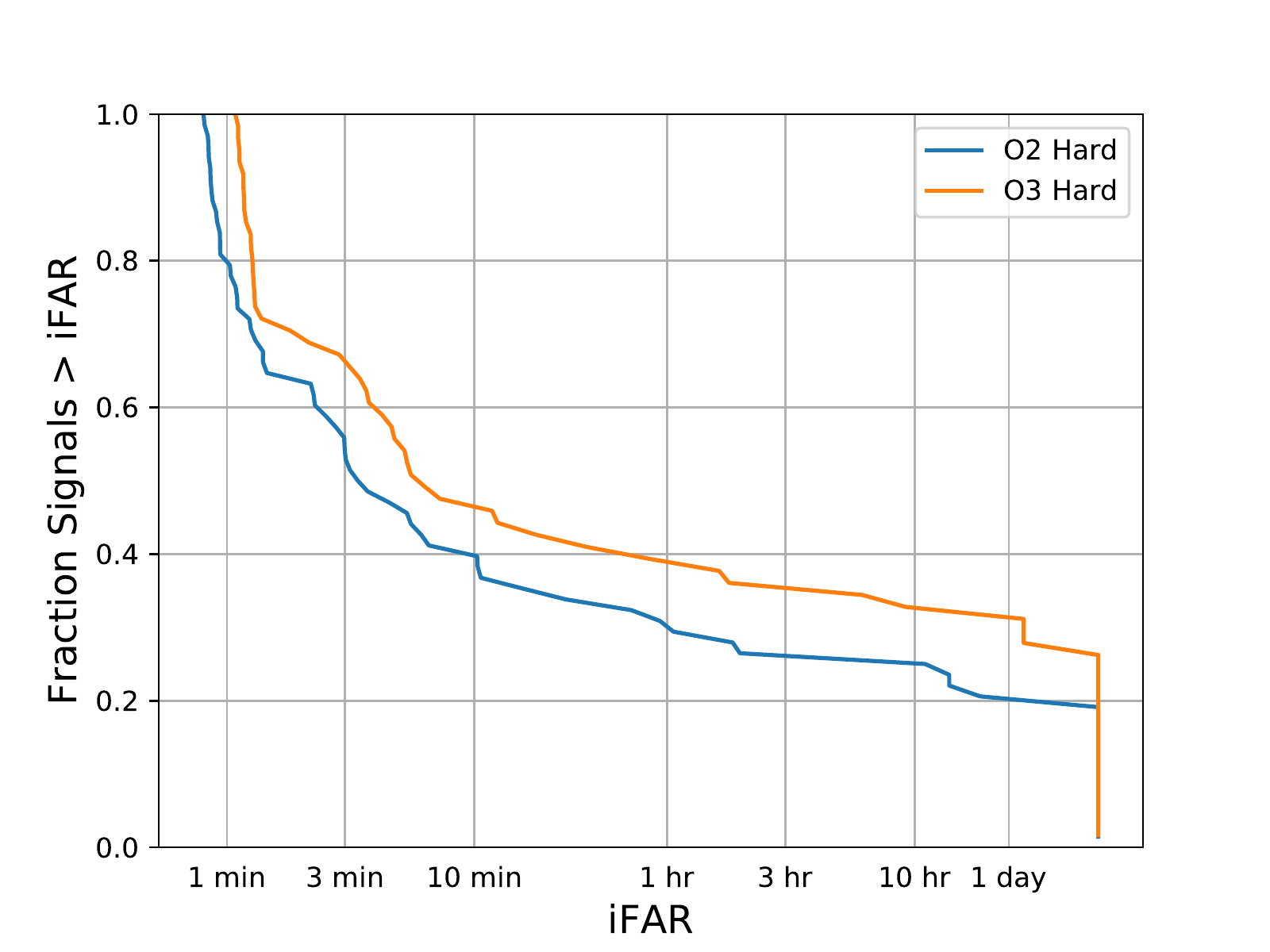}
		\includegraphics[scale=0.5]{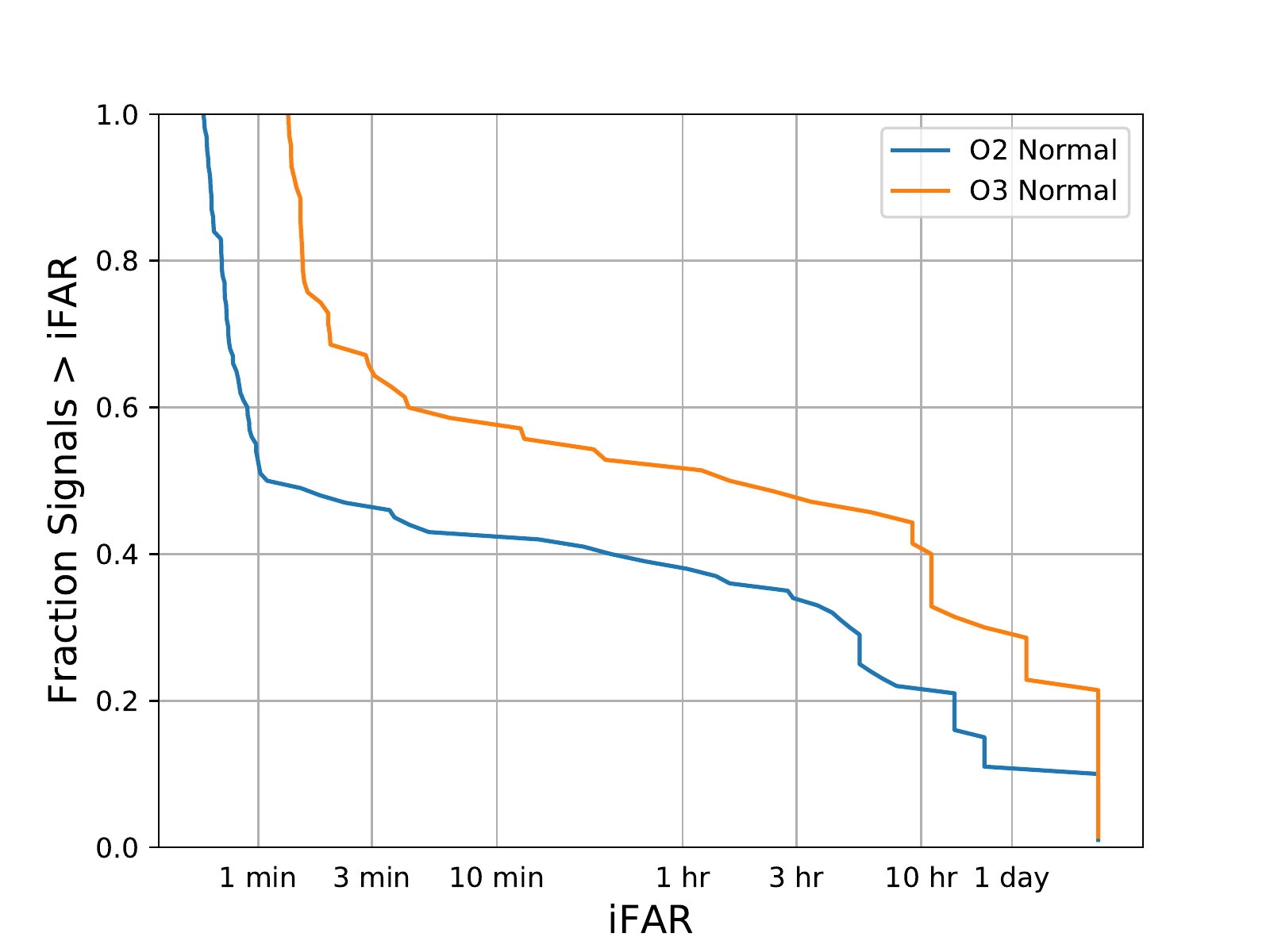}\\
		\includegraphics[scale=0.5]{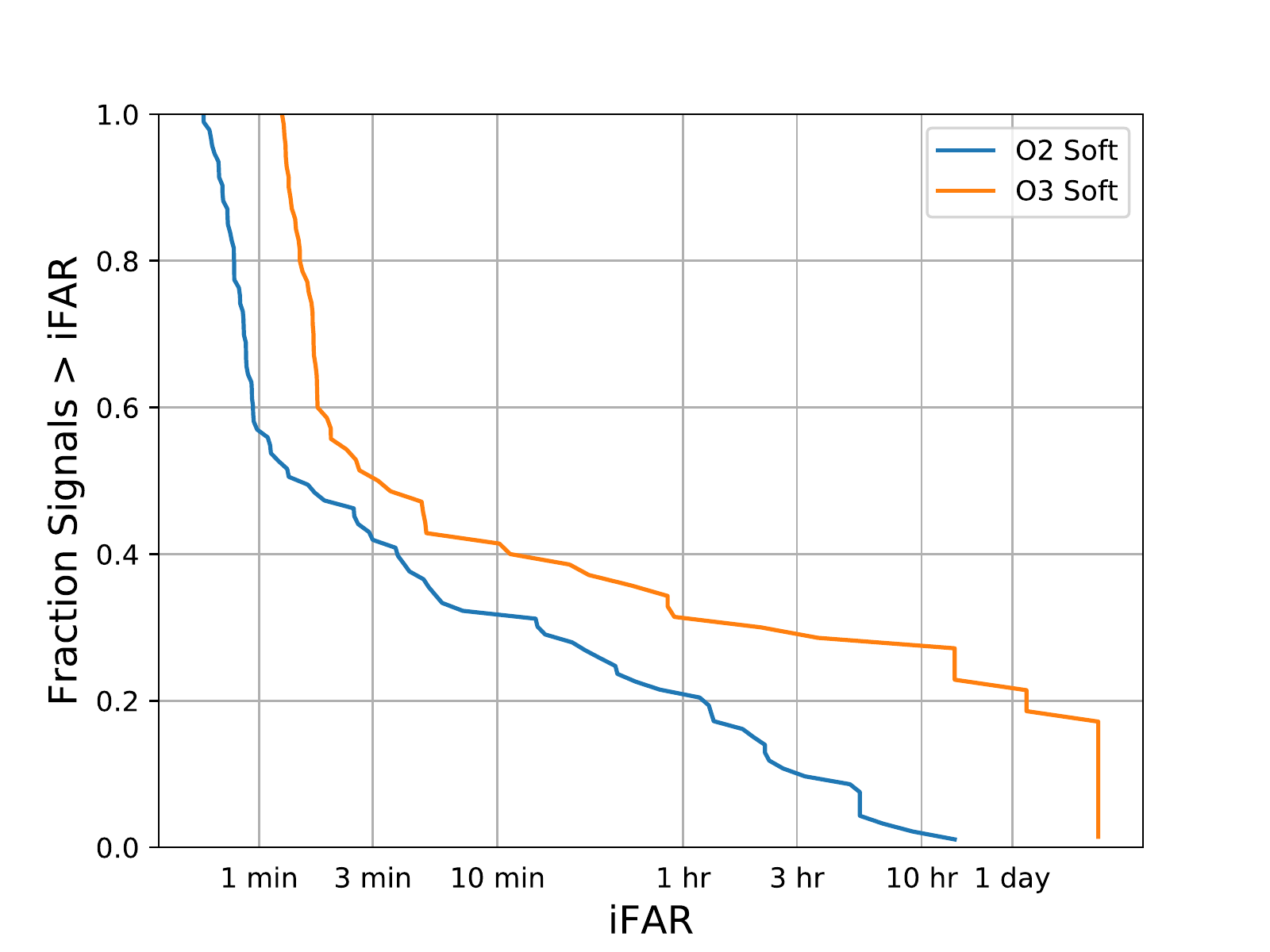}\\
	\end{center}
\caption{Cumulative distributions of real signal significance compared between the O2 and O3 searches, plotted as inverse FAR.  Consistently, a larger fraction of real signals are discovered at a larger significance (lower FAR) for all templates in the O3 search.
\label{FinalSig}}
\end{figure}

\begin{figure}
	\begin{center}
		\includegraphics[scale=0.6]{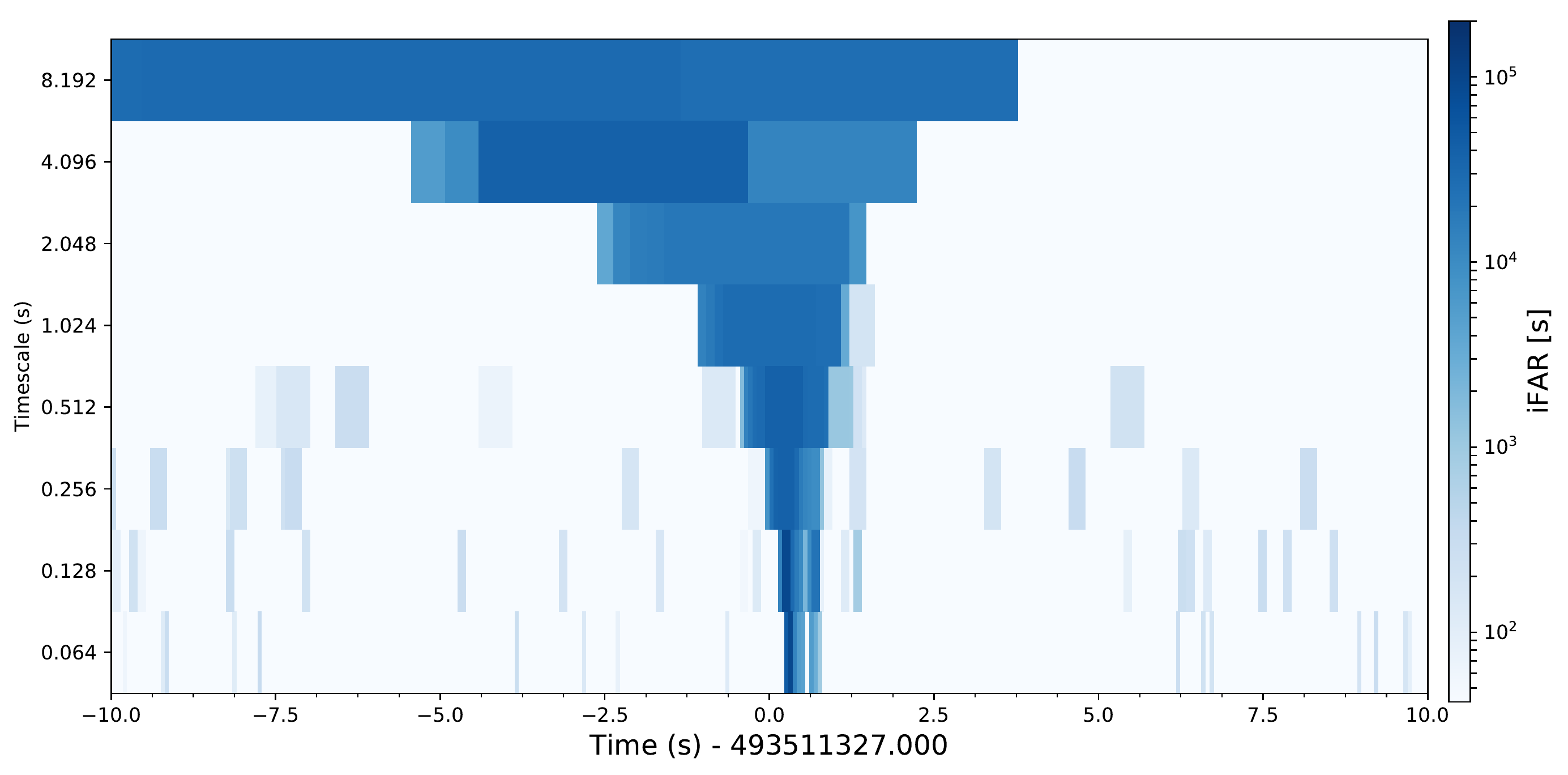}\\
		\includegraphics[scale=0.6]{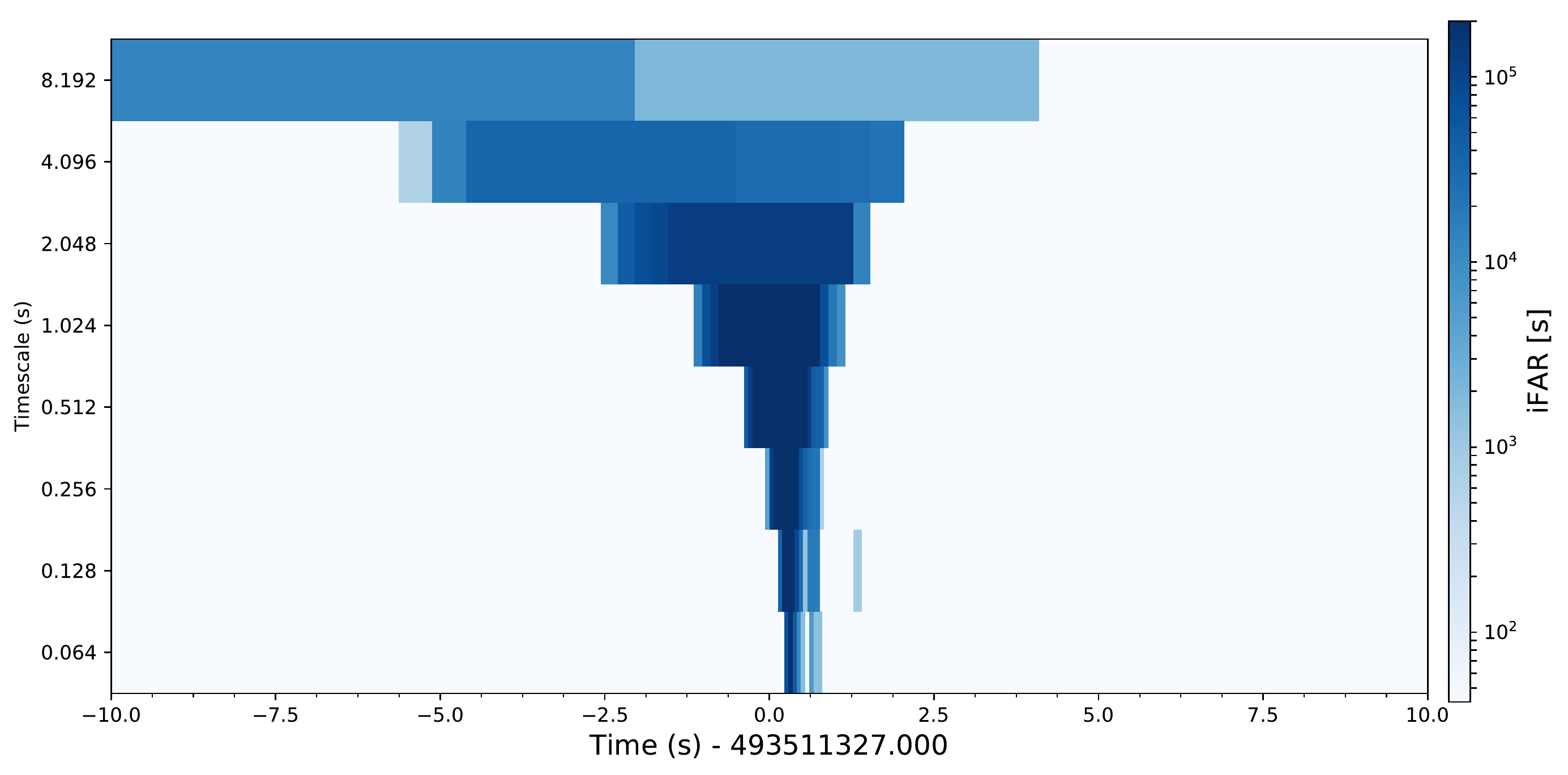}
	\end{center}
\caption{Waterfall plot comparison between the O2 (top) and O3 (bottom) searches for GRB 160821B.  The color gradient represents the inverse FAR.  The noticeable difference is the increased significance of the GRB peak in the O3 search.
\label{DontGoChasingWaterfalls}}
\end{figure}

The significance of the detection statistic (LogLR) for a potential candidate signal is determined by the FAR distribution generated by the background.  Both are dependent on the characteristics of the search, therefore to evaluate the overall performance of the search to find real signals, a candidate's LogLR must be compared to the background distribution.  In Figure~\ref{FinalSig} we show the cumulative distributions of the inverse FAR (iFAR) of the signal validation samples. An improvement to the sensitivity of a search will exhibit an increased cumulative fraction shifted toward higher iFAR.  The improvement in sensitivity in the O3 search compared to the O2 seach is clear; the O3 search is more sensitive in all three spectral templates from an iFAR of 1 minute up to, and exceeding, an iFAR of 1 day.  An example of how this improvement manifests during a search is shown in Figure~\ref{DontGoChasingWaterfalls}, which displays the iFAR for each timescale and phase shift searched during the window of interest.

Significant improvements beyond an iFAR of $\sim1$ day using only the temporal information of the search will be difficult since the GBM onboard trigger rate of GRBs is about 1 per 1.5 days.  Further improvement to the iFAR can be made if the localization information can be incorporated into the statistic.\\

\section{Upper Limits\label{sec:UpperLimits}}
The targeted search detection statistic is a log-likelihood ratio comparison of the alternative hypothesis of the presence of a signal plus background noise to the null hypothesis of only the presence of noise. \citet{Blackburn15} formulated an amplitude-marginalized log-likelihood ratio that requires the numerical estimate of the signal amplitude that maximizes the likelihood:

\begin{equation}
	\mathcal{L}(d) = \ln{\sigma_\mathcal{L}} + \ln{\biggl[1 + {\rm Erf}\biggl(\frac{s_{\rm max}}{\sqrt{2}\sigma_ \mathcal{L}}\biggr)\biggr]} +  \mathcal{L}(d|s_{\rm max}) + 
	\begin{cases}
  		\ln{\bigl[1-e^{-(s_{\rm max}/\gamma \sigma_\mathcal{L})}}\bigr]-\ln{s_{\rm max}} & s_{\rm max } > 0\\
		-\ln{(\gamma \sigma_\mathcal{L}}) & s_{\rm max} \leq 0
	\end{cases}
\end{equation}
where $s_{\rm max}$ is the signal amplitude (photon flux) that maximizes the likelihood, $\sigma^2_\mathcal{L}$ is the variance of the likelihood assuming $s_{\rm max}$, $\mathcal{L}(d|s_{\rm max})$ is the likelihood of the data assuming $s_{\rm max}$, and the final term results from utilizing a power-law prior for the amplitude.  This equation is evaluated for each bin of data, assuming a background model, and utilizing the detector responses for each point on the visible sky grid. By integrating over the sky, $\mathcal{L}(d)$ is the detection statistic, LogLR, reported by the targeted search.  For each bin of data, we calculate $s_{\rm max}$ and $\sigma_\mathcal{L}$, which represents the 1-sigma uncertainty in the photon flux.  Assuming a Gaussian distribution of the uncertainty, we can calculate a $3-\sigma$ photon flux upper limit, $S_{\rm UL}$:
\begin{equation}
	S_{\rm UL} = s_{\rm max} + 3\sigma_\mathcal{L}.
\end{equation}
This upper limit can be calculated for each visible point in the sky grid to create an upper limit map on the visible sky.  Because we search over a defined span of time and on different timescales, it is appropriate to quote different upper limits based on different timescales of the search, and the most conservative upper limit for each timescale is determined by taking the maximum upper limit over the duration of the search window.  Furthermore, if the targeted search is given an external sky map (e.g. a GW localization map), it can be treated as a spatial prior for the upper limits calculation.  To do this, we weight the photon flux upper limit for each sky point by the spatial prior probability at that point and marginalize across the spatial prior:
\begin{equation}
	\bar{S}_{\rm UL} = \sum_i^{n_{\rm sky}}{P_{{\rm spatial},i} \ S_{{\rm UL},i}}.
\end{equation}
Finally, using the assumed spectral templates, the photon flux upper limit can be converted to an energy flux (or fluence) upper limit for each timescale.  We use this method beginning in O3a to calculate the preliminary upper limits for the LVC open public alerts.

\section{Localization Systematic\label{sec:localization}}
\begin{figure}
	\begin{center}
		\includegraphics[scale=0.55]{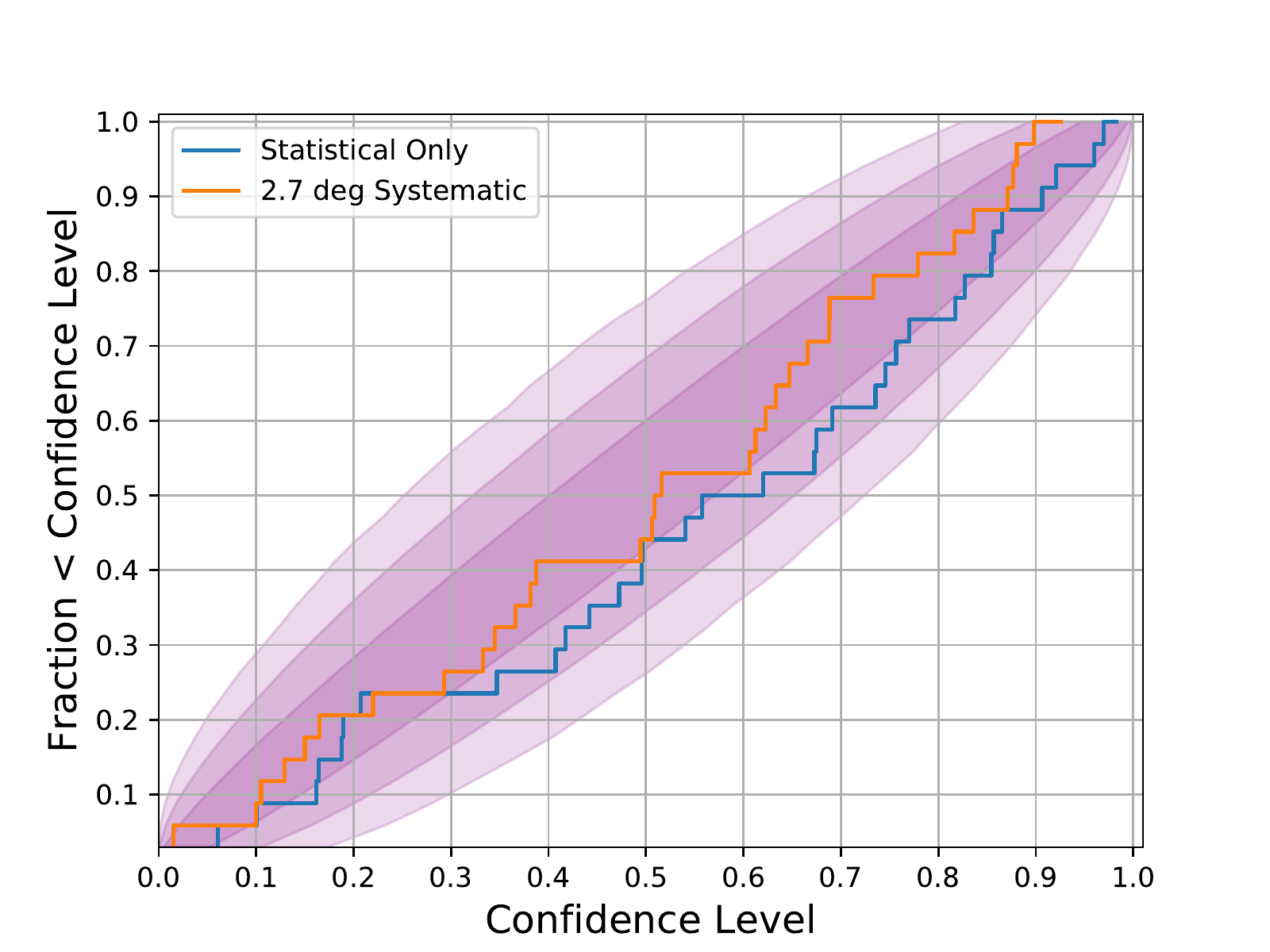}
		\includegraphics[scale=0.55]{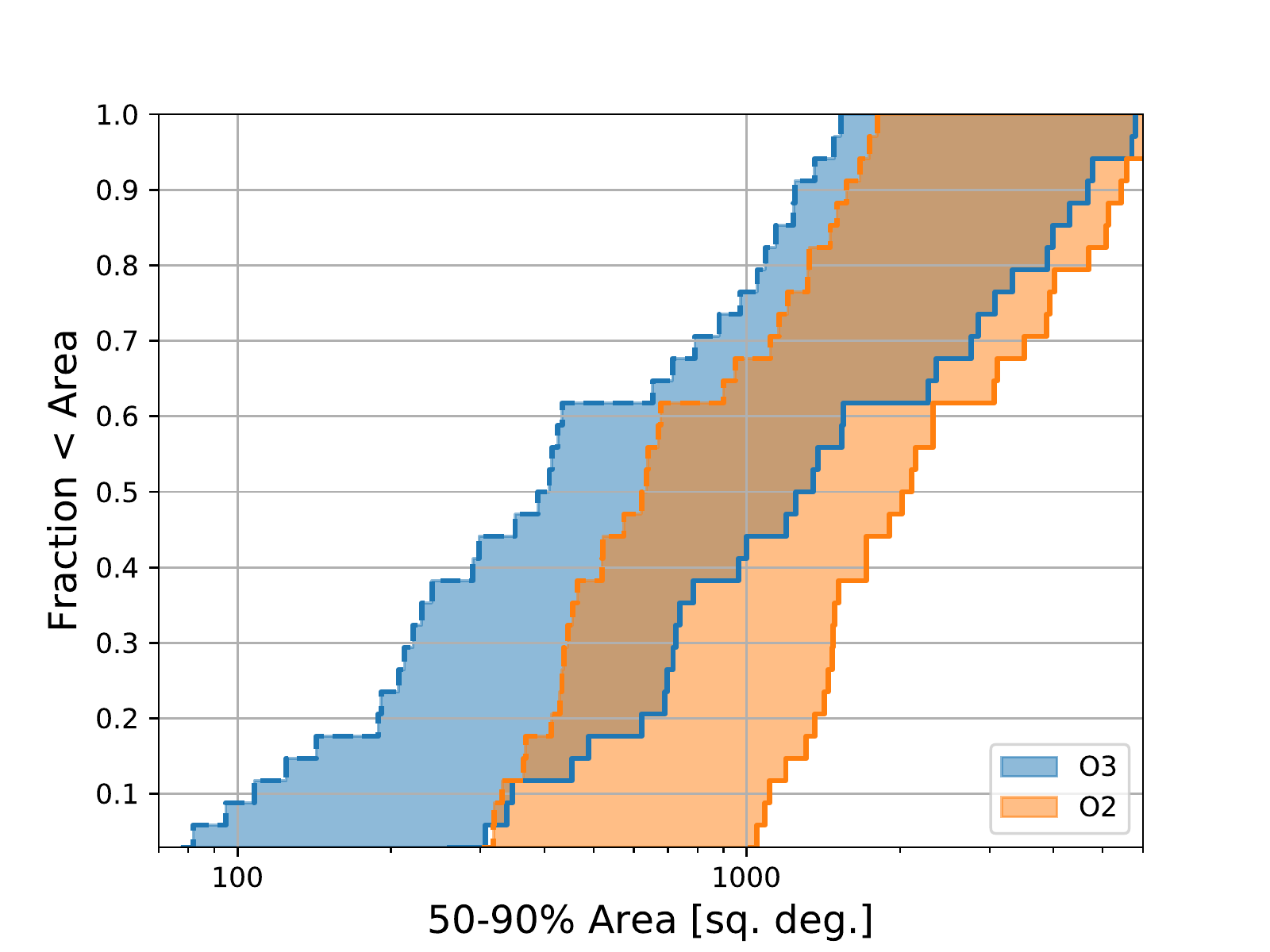}
	\end{center}
\caption{(Left) The probability-probability plot of the localization from the targeted search for short GRBs with known location. The purple bands denote the 1, 2, and 3 $\sigma$ confidence regions for the calibration based on the sample of 34 localizations.  A $\sim2.7^\circ$ systematic is required.  (Right) The comparison of the cumulative distributions of the 50\% and 90\% localization areas between the systematic used in O2 ($7.6^\circ$) and the O3 systematic.
\label{LocSystematic}}
\end{figure}

Because the targeted search calculates the LogLR over the visible sky, a likelihood map on the sky can be created and converted to a localization posterior.  This posterior incorporates the statistical uncertainty, and therefore may be an underestimate of the true uncertainty.  For the O2 search, the localization systematic assessment was performed, prompted by the joint detection of GW170817/GRB~170817A in order to estimate the joint spatial coincidence.  That study utilized short GRBs that triggered onboard the spacecraft to estimate the systematic in the O2 targeted search: $7.6^\circ$~\citep{Goldstein17}.  For the O3 search, we utilized sub-threshold short GRBs discovered by the GBM untargeted blind search that were also detect by Swift, in addition to sub-threshold GRBs contained within~\citet{Kocevski18}, providing a sample of 34 GRBs.  This modest sample allows us to estimate the localization systematic of the O3 targeted search following the procedure in~\citet{Goldstein19}: maximizing the binomial likelihood of finding the true location of the sub-threshold GRBs within the containment region defined by the targeted search localization posterior.  Figure~\ref{LocSystematic} shows the probability-probability calibration plot of this sample without a systematic uncertainty compared to the maximum likelihood fit of a Gaussian systematic component.  In general, a $2.7^\circ$ systematic uncertainty is required, which is significantly smaller than what was required in the O2 version of the search.  Figure~\ref{LocSystematic} also shows the comparison of the final posterior areas once the systematic is incorporated for the two versions of the search.  Note that this is a limited sample and may not represent the full distribution of possible sub-threshold localizations, however the area reduction is significant, with $\sim57$\% median reduction in area for both the 50\% and 90\% credible regions.

Further improvement of the targeted search localization can be attained by operating the targeted search in a signal characterization mode once a candidate has been found.  This mode operates the search on a smaller window centered on the candidate time, covering finer timescales and a larger number of binning phase shifts of the data in order to optimize the model signal-to-noise and to produce and optimal localization.  In this mode, the localization from the targeted search becomes $0.5^\circ$ more accurate on average and the localization area reduces by another 27\% on average.  This mode is only invoked if a promising candidate is found, such as was the case for Fermi-GBM 190816~\citep{190816_v1, 190816_v2}.

\section{Case Study: Fermi GBM-190816\label{sec:190816}}
\begin{figure}
	\begin{center}
		\includegraphics[scale=0.5]{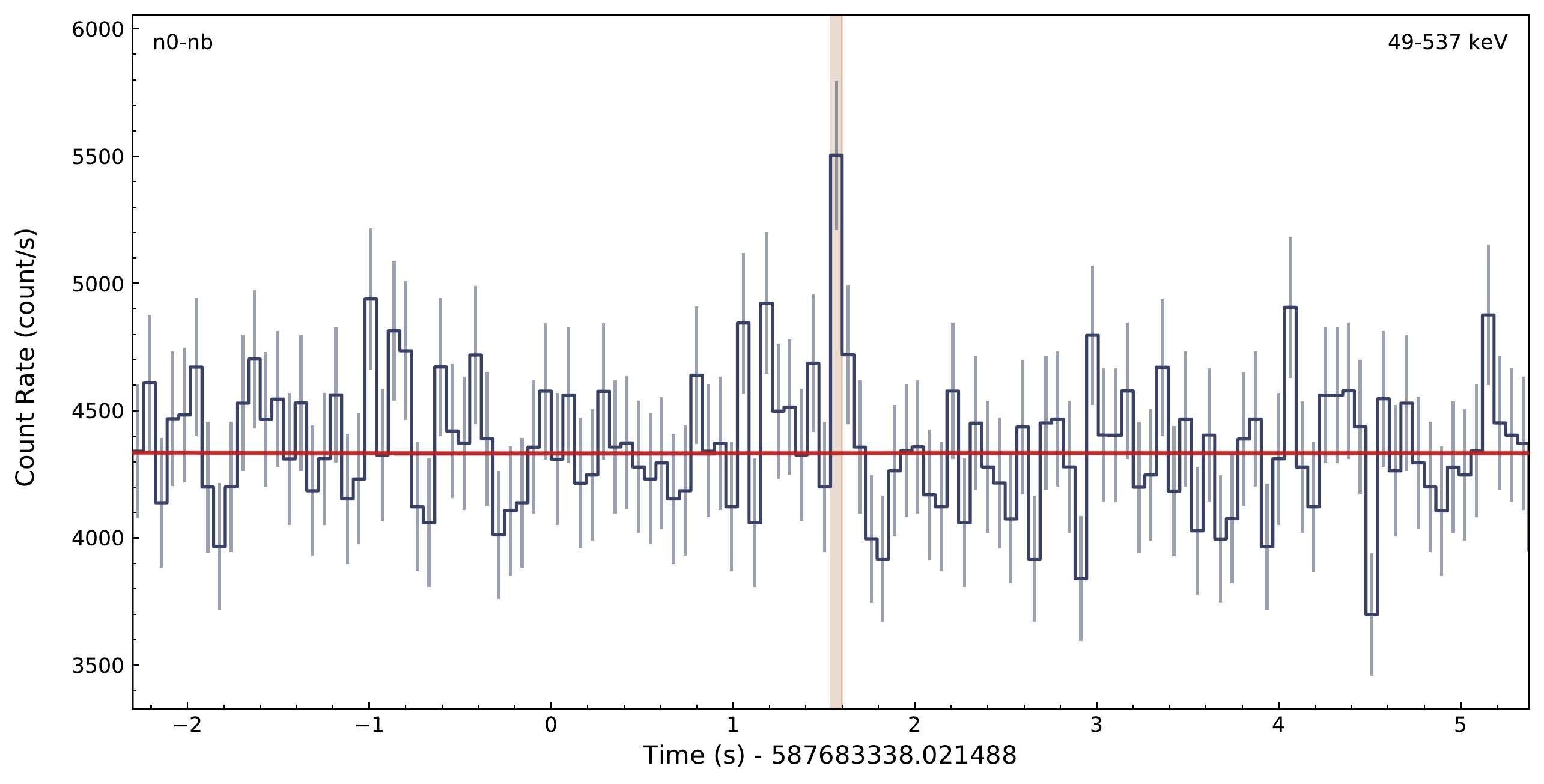}\\
		\includegraphics[scale=0.5]{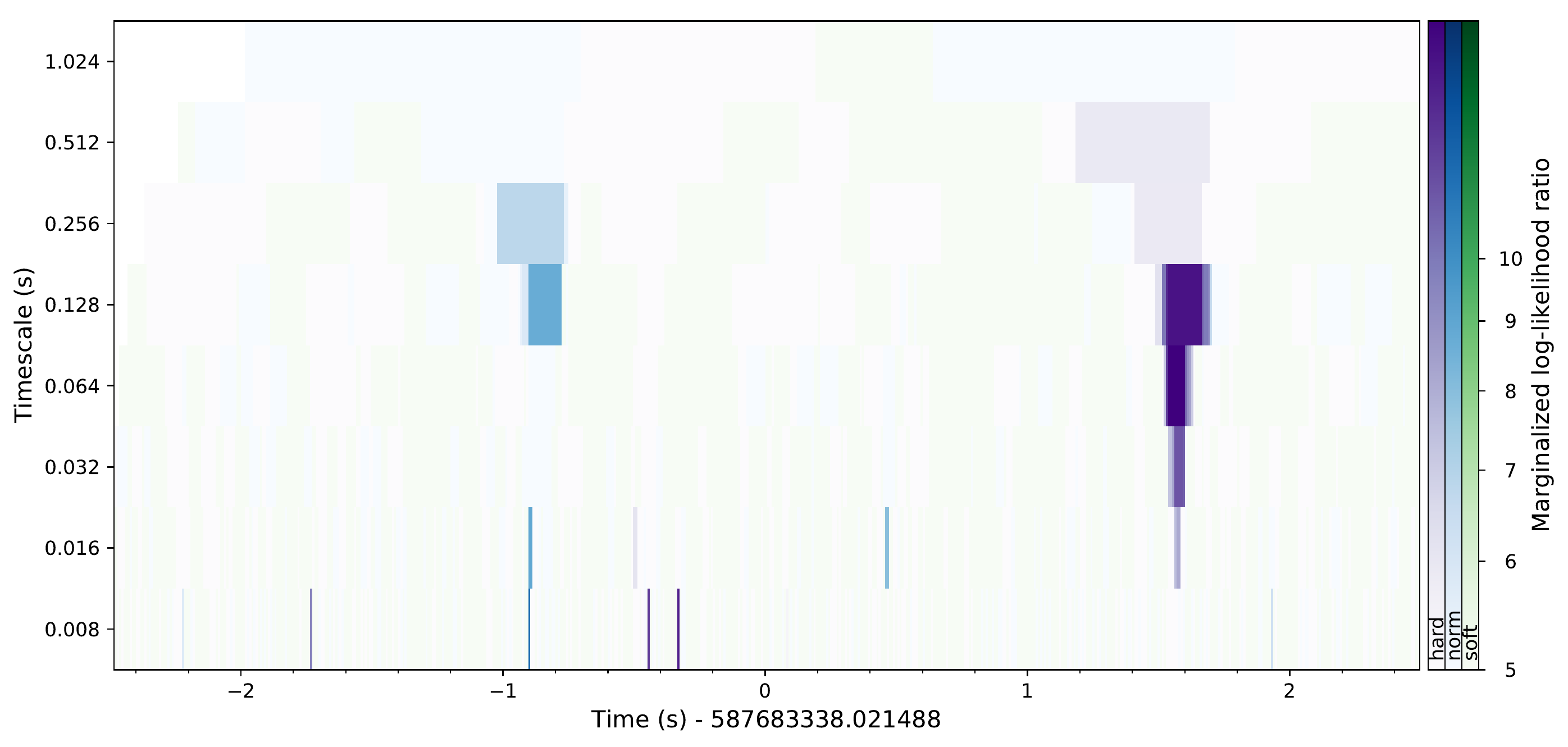}\\
		\includegraphics[scale=0.65]{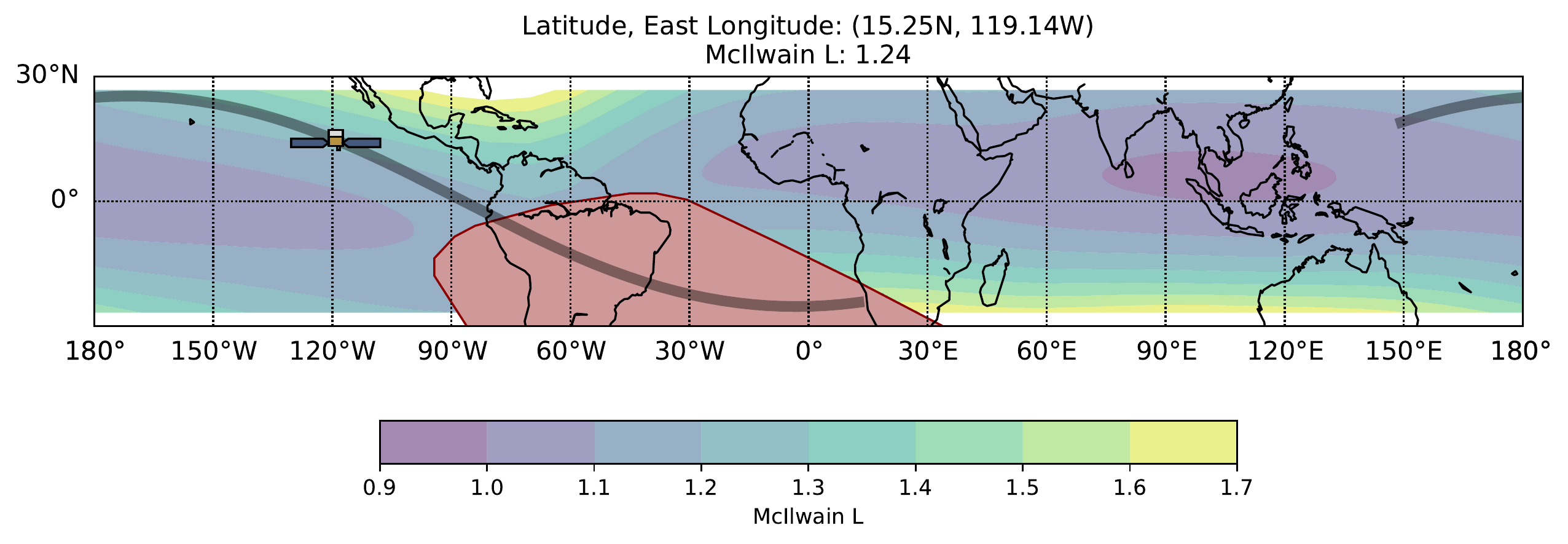}
	\end{center}
\caption{The sub-threshold short GRB candidate Fermi GBM-190816, found $\sim1.6$ s after a sub-threshold GW candidate. The lightcurve (top) shown on the 64 ms timescale relative to the GW trigger time, summed over all NaI detectors.  The waterfall plot (middle) on finer timescales and divided into the different spectral templates to aid in characterization of the signal.  (Bottom) Location of the Fermi spaceraft in orbit at the time of GBM-190816.  The gray shaded line is the path of the orbit and the color gradient represents the geomagnetic latitude in terms of the McIlwain L.  If the spacecraft is in an area where McIlwain L is $\gtrsim 1.5$, the chance that a trigger is related to local particles is significant.
\label{GBM-190816}}
\end{figure}

As an example of the capability of the targeted search during O3, we present the case of the sub-threshold gamma-ray transient Fermi GBM-190816~\citep{190816_v1}.  This short GRB candidate was found during routine follow-up of sub-threshold gravitational-wave triggers from LIGO and Virgo, and one key property that makes it especially interesting is the fact that it occurred $\sim1.6$ s after the GW trigger.  Figure~\ref{GBM-190816} shows the lightcurve, summed over all NaI detectors in the energy range of $\sim50-500$ keV.  The targeted search found this candidate with the hard spectral template on a timescale of 64 ms at a LogLR=14.9.  The FAR, irrespective of the GW signal significance, is about $1.2\times10^{-4}$ Hz, or about 1 false alarm per 2.3 hours searched.  To further characterize the signal, especially since it was found on the shortest timescale of the search, we reran the search down to 8 ms and with a maximum of 16 phase steps (e.g. each bin in the 256 ms timescale is shifted by 16 ms, 16 times).  Figure~\ref{GBM-190816} shows the waterfall plot of this characterization, and the signal can be seen on multiple timescales, spanning $\sim16-128$ ms.  The structure of this signal over several consecutive timescales is favorable for a coherent signal with a pulse-like shape and is similar to those found for known sub-threshold short GRBs~\citep{Kocevski18}, and is unlike the structure of a statistical fluctuation.

To further characterize the signal in GBM, several factors are important to consider.  The location of Fermi in orbit is important because the orbit crosses geomagnetic latitudes that are conducive to the detection of particles trapped in the magnetosphere.  Similarly, Fermi's orbit takes it through the South Atlantic Anomaly which can rapidly affect the background noise rate on approach or exit, leading to an increased chance in a poor background fit.  At the time of the detection, Fermi was not near the SAA and was at a favorable geomagnetic latitude.  The chance of local particle activity was minimal, and the apparent spectrum and duration of the signal was not favorable for a local particle source.  Additionally, while the orbital position of the spacecraft was near the edge of a favorable location (off the west coast of Mexico) for the detection of Terrestrial Gamma-ray Flashes (TGFs), the signal duration is orders of magnitude longer than the duration of TGFs (typically microseconds timescale).  Considering the localization of the event, it was somewhat consistent with part of the Galactic plane near the anti-center, and while GBM detects many Galactic sources while they are active, Galactic sources are typically quite spectrally soft and long in duration, inconsistent with the observed signal characteristics of GBM-190816.  Finally, GBM often detects solar flare activity from the sun that can appear on a wide variety of timescales.  While the refined localization did move toward the location of the sun from the initial localization, GOES solar flux monitoring data showed no solar activity that deviated from the baseline activity at the time of the event, and the hard spectrum of the event is inconsistent with the typically much softer spectrum observed by GBM for solar flares.  Considering all of the evidence from the characterization, the gamma-ray transient signal Fermi GBM-190816 is most likely a short GRB, however due to its sub-threshold nature, a solid claim of the detection of a short GRB would require confirmation from another instrument.

\section*{Acknowledgment}
\noindent
The improvement and operations of the {\it Fermi} GBM O3 targeted sub-threshold search is supported by the {\it Fermi} Guest Investigator Program (NNH17ZDA001N-FERMI).


\clearpage

\end{document}